\documentclass[%
 reprint,
 superscriptaddress,
 amsmath,amssymb,
 aps,
prb,
floatfix,
]{revtex4-2}

\usepackage{graphicx}
\usepackage{dcolumn}
\usepackage{bm}

\newcommand{\picWidth}{8.5cm}
\newcommand{\widePicWidth}{17cm}

\usepackage{hyperref}
\hypersetup{colorlinks,allcolors=blue}

\begin{document}

\title{The Influence of Extended Interactions on Spin Dynamics in One-dimensional Cuprates}

\author{Ta Tang}
\email{tatang@stanford.edu}
\affiliation{Stanford Institute for Materials and Energy Sciences,
SLAC National Accelerator Laboratory, 2575 Sand Hill Road, Menlo Park, CA 94025, USA}
\affiliation{Department of Applied Physics, Stanford University, Stanford, CA 94305, USA}

\author{Daniel Jost}
\affiliation{Stanford Institute for Materials and Energy Sciences,
SLAC National Accelerator Laboratory, 2575 Sand Hill Road, Menlo Park, CA 94025, USA}

\author{Brian Moritz}
\affiliation{Stanford Institute for Materials and Energy Sciences,
SLAC National Accelerator Laboratory, 2575 Sand Hill Road, Menlo Park, CA 94025, USA}

\author{Thomas P. Devereaux}
\email{tpd@stanford.edu}
\affiliation{Stanford Institute for Materials and Energy Sciences,
SLAC National Accelerator Laboratory, 2575 Sand Hill Road, Menlo Park, CA 94025, USA}
\affiliation{Department of Materials Science and Engineering, Stanford University, Stanford, CA 94305, USA}
\affiliation{Geballe Laboratory for Advanced Materials, Stanford University, Stanford, California 94305, USA}

\date{\today}

\begin{abstract}
Quasi-one-dimensional (1D) materials provide a unique platform for understanding the importance and influence of extended interactions on the physics of strongly correlated systems due to their relative structural simplicity and the existence of powerful theoretical tools well-adapted to one spatial dimension. Recently, this was highlighted by anomalous observations in the single-particle spectral function $A(q,\omega)$ of 1D cuprate chain compounds, measured by angle-resolved photoemission spectroscopy (ARPES), which were explained by the presence of a long-range attractive interaction. Such an extended interaction should leave its fingerprints on other observables, notably the dynamical spin structure factor $S(q,\omega)$, measured by neutron scattering or resonant inelastic x-ray scattering (RIXS). Starting from a simple Hubbard Hamiltonian in 1D and using time-dependent density matrix renormalization group (tDMRG) methods, we show that the presence of long-range attractive coupling, directly through an instantaneous Coulomb interaction $V$ or retarded electron-phonon ({\it el-ph}) coupling, can introduce significant spectral weight redistribution in $S(q,\omega)$ across a wide range of doping. This underscores the significant impact that extended interactions can have on dynamical correlations among particles, and the importance of properly incorporating this influence in modeling. Our results demonstrate that $S(q,\omega)$ can provide a sensitive experimental constraint, which complements ARPES measurements, in identifying key interactions in 1D cuprates, beyond the standard Hubbard model. 
\end{abstract} 

\maketitle

The origin of high-temperature superconductivity that has been found in layered, quasi-two-dimensional (2D) cuprates remains elusive despite concerted investigations over the last few decades. 
From the perspective of numerical simulations, although simplified Hamiltonians, such as the Hubbard model and related variants, have produced rich physics, seemingly relevant to the cuprates \cite{dagottoCorrelatedElectrons1994, arovasHubbardModel2022,qinHubbardModel2022}, insufficient evidence exists for the presence of $d$-wave superconductivity in the ground state. 
Quasi-long-range superconductivity has been reported only on small width cylinders with a strong competition from coexisting charge order~\cite{whiteGroundStates1997,ehlersHybridspaceDensity2017,simonscollaborationonthemany-electronproblemSolutionsTwoDimensional2015,jiangSuperconductivityDopedHubbard2019c,thesimonscollaborationonthemany-electronproblemPlaquetteOrdinary2020,jiangGroundStatePhase2020,jiangGroundstatePhaseDiagram2021,gongRobustWave2021,jiangPairingPropertiesModel2022,simonscollaborationonthemany-electronproblemAbsenceSuperconductivityPure2020}. 
Moreover, it seems superconducting correlations decay exponentially for wider clusters, indicating that superconductivity may be absent for parameters thought to be relevant to the hole-doped cuprate compounds. 

While numerical difficulties in reaching the 2D limit may prevent us from saying anything definitive about superconductivity in these simplified models, additional ingredients may well be needed to provide the crucial boost for superconductivity. 
Assessing the impact of missing ingredients, such as phonons, whose influence has manifest as kinks or replica bands in photoemission measurements~\cite{lanzaraEvidenceUbiquitousStrong2001, cukReviewElectron2005, leeInterplayElectronLattice2006, heRapidChangeSuperconductivity2018, kulicInterplayElectron2000}, or long-range interactions, is even more challenging in 2D modeling due to the additional numerical difficulty associated with adding more degrees of freedom and the associated expansion of the effective Hilbert space, and the growth of long-range entanglement. Our task may be made easier, with more numerical control and better theoretical understanding, by turning to the simpler, yet structurally similar, quasi-one-dimensional (1D) cuprates.

\begin{figure*}[th]
    \begin{center}
        \includegraphics[width=\widePicWidth]{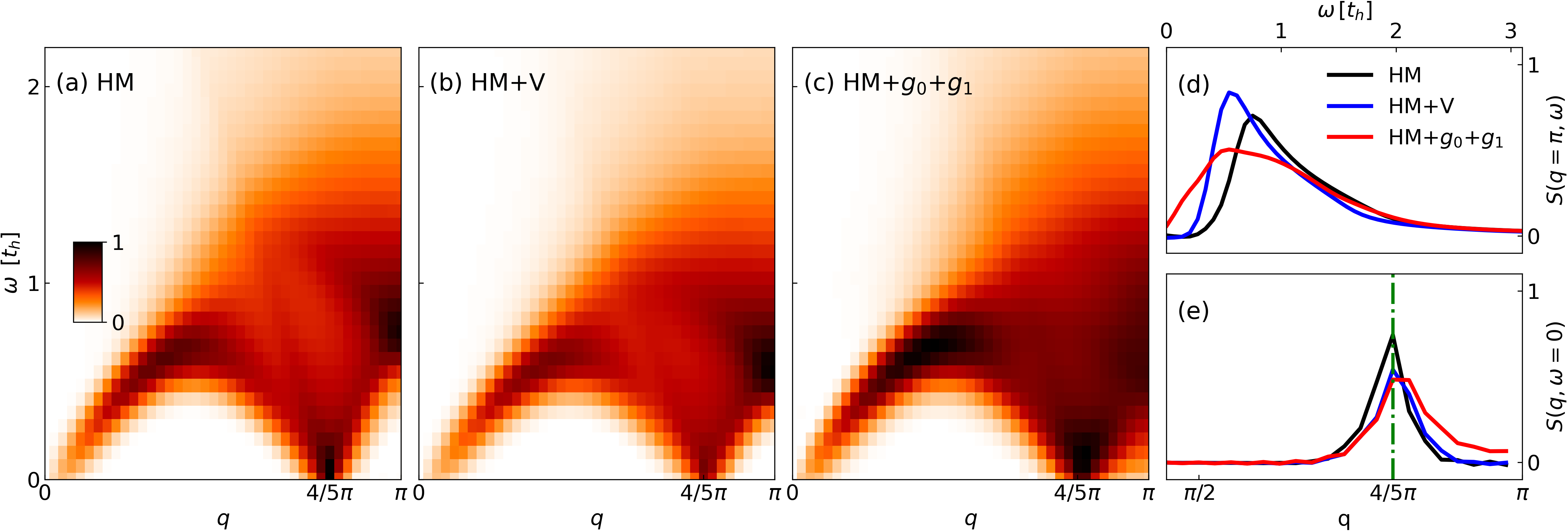} 
    \end{center}
    \caption{
        \textbf{Dynamical spin structure factor with L=80.}
         $S(q,\omega)$ from different models at 20\% doping on an 80-site chain with broadening $\delta=0.1t_h$. 
         See Fig.~\ref{pic:L80_srt} in the Supplementary Material for the real-space, time-dependent spin-spin correlations. 
         Compared to the HM in (a) and HM+$V$ in (b), excitations in $S(q,\omega)$ for the HM+$g_{0}$+$g_{1}$ in (c) are much broader and appear to be gapless across momentum $q_0$ to $\pi$. 
         (d) EDC at $q=\pi$. Extended {\it el-ph} coupling shifts the peak towards $\omega=0$ and broadens the spectra. (e) MDC at $\omega=0$. Green dashed line indicate $q_0=(1-\delta)\pi = 4/5\pi$. While the peaks of both the HM and HM+$V$ are located at $q_0$, the extended {\it el-ph} coupling broadens and shifts the peak from $q_0$ towards $q=\pi$.
     }
    \label{pic:L80_sqw}
\end{figure*}

Recent synthesis progress on the doped 1D-chain cuprate $\mathrm{Ba}_{2-x}\mathrm{Sr}_x\mathrm{CuO}_{3+\delta}$~\cite{chenAnomalouslyStrongNearneighbor2021} provides an excellent opportunity for testing theoretical models against experiments. 
Angle-resolved photoemission spectroscopy (ARPES) data on this doped 1D-chain in conjunction with numerical simulations revealed the presence of a strong, extended, attractive interaction~\cite{chenAnomalouslyStrongNearneighbor2021}. Such an effective strong and extended attractive interaction likely originates from extended electron-phonon ({\it el-ph}) couplings~\cite{wangPhononMediatedLongRangeAttractive2021}. It was found in later numerical work~\cite{tangTracesElectronphonon2023}, that the extended {\it el-ph} coupling reproduces well the doping dependence of salient ARPES spectral features, which has been shown to enhance superconducting pairing correlations in 1D and have important implications for realizing a $d$-wave superconducting ground state in the structurally similar layered 2D cuprates. This extended {\it el-ph} coupling can significantly suppress the static spin-spin correlation~\cite{tangTracesElectronphonon2023} otherwise obtained from the Hubbard model, while only slightly affecting the exponent of the single particle correlations. 
This motivates us to unravel the spin-spin correlation in frequency through the dynamical spin structure factor $S(q,\omega)$, which can be used to further refine how the {\it el-ph} interactions impact the spin dynamics beyond the simple Hubbard model. As we will show, $S(q,\omega)$ may provide a sensitive probe on the presence and influence of extended {\it el-ph} interactions in the 1D cuprates.

In this work, we compute the dynamical spin structure factor $S(q,\omega)$ influenced by extended, attractive couplings on a 1D chain for various hole doping concentrations $\delta$. We demonstrate that the addition of such extended interactions can introduce significant qualitative and quantitative changes to $S(q,\omega)$ as a function of $\delta$. The two-spinon excitations at the zone edge significantly broaden and soften, even becoming gapless across the momentum range $(1-\delta)\pi \le q \le \pi$ when the extended interactions come from {\it el-ph} coupling, while also leading to an apparent shift in the ``2$k_{\textrm{F}}$" momentum position towards $\pi$, compared to the simple Hubbard model [$2k_{\textrm{F}}=(1-\delta)\pi$] .

The Hubbard model (HM) is purely electronic
\begin{equation}
H_{el} = -t_h\sum_{\left<ij\right>\sigma}(\hat{c}^\dagger_{i\sigma}\hat{c}_{j\sigma} + h.c.) + U\sum_{i}\hat{n}_{i\uparrow}\hat{n}_{i\downarrow},
\label{eq:hubbard}
\end{equation}
where $\hat{c}^\dagger_{i\sigma}$ ($\hat{c}_{i\sigma}$) is the charge creation (annihilation) operator on site $i$ for spin $\sigma$, $\hat{n}_{i\sigma}$ is the charge number operator on site $i$ for spin $\sigma$, and $U$ is the on-site repulsion. To avoid confusion with the time variable $t$, we use $t_h$ to denote the hopping integral. We consider two forms of extended, attractive couplings to modify this simple HM. In one case, we introduce an effective nearest-neighbor attractive interaction, which has been used in prior work to well-reproduce the apparent single-particle spectra from ARPES experiments~\cite{chenAnomalouslyStrongNearneighbor2021}, to produce an extended-Hubbard model (HM+$V$) given by
\begin{equation}
H_{v} = H_{el} + V\sum_{\left<ij\right>}\hat{n}_i\hat{n}_j,
\label{eq:extended-hubbard}
\end{equation}
where $\hat{n}_i$ and $\hat{n}_j$ are total charge number operators on neighboring sites. In the other case, we add local optical phonons with coupling to the on-site and nearest-neighbor charge densities to the HM, resulting in a Hubbard-extended Holstein model (HM+$g_{0}$+$g_{1}$)~\cite{wangPhononMediatedLongRangeAttractive2021, tangTracesElectronphonon2023}
\begin{eqnarray}
    H &=& H_{el} + \omega_0\sum_i \hat{a}^\dagger_i\hat{a}_i \nonumber \\
    &&\quad + g_0\sum_i \hat{n}_i (\hat{a}^\dagger_i + \hat{a}_i) + g_1\sum_{\left<ij\right>} \hat{n}_i (\hat{a}^\dagger_j + \hat{a}_j),
    \label{eq:hhm}
\end{eqnarray}
where $\hat{a}^\dagger_i$ and $\hat{a}_i$ are the phonon ladder operators on site $i$, $\hat{n}_i$ is the total charge number operator on site $i$, $\omega_0$ is the phonon frequency, $g_0$ is the on-site {\it el-ph} coupling, $g_1$ is the nearest-neighbor {\it el-ph} coupling, and $\left<ij\right>$ sums over nearest-neighbors.

\begin{figure*}[th]
    \begin{center}
        \includegraphics[width=\widePicWidth]{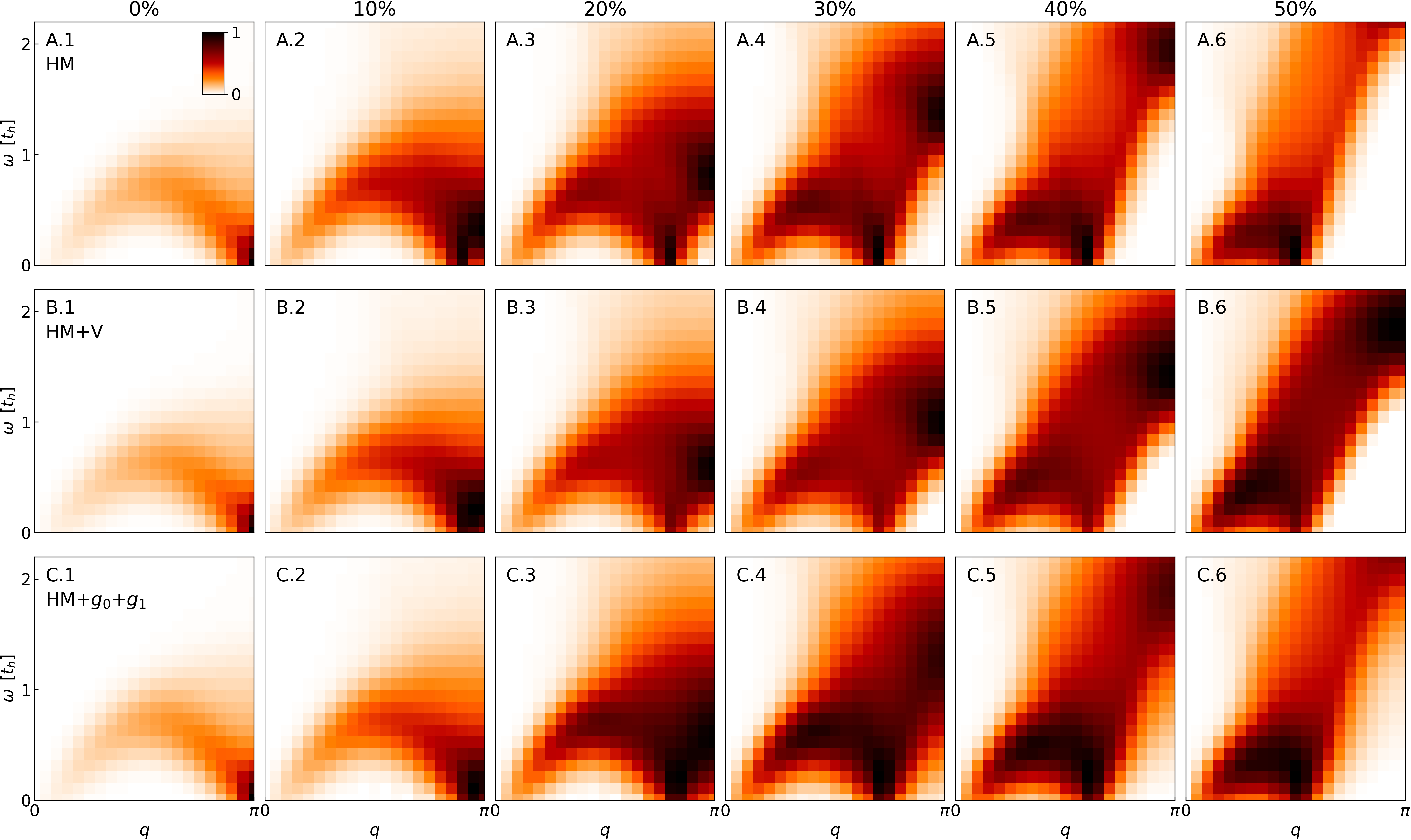} 
    \end{center}
    \caption{
        \textbf{Dynamical spin structure factor with L=40.}
         $S(q,\omega)$ for  model Hamiltonians at doping concentrations as indicated on a 40-site chain with broadening $\eta=0.2t_h$. Row A is for the Hubbard model (HM), row B is for the extended-Hubbard model (HM+$V$), and row C is for the Hubbard-extended Holstein model (HM+$g_{0}$+$g_{1}$). Each of the spectra have been normalized to their separate maxima. Columns 1-6 correspond to 0\%-50\% doping. 
         $S(q,\omega)$ has been obtained from the real-space, time-dependent spin-spin correlation (see Fig.~\ref{pic:L40_srt} in the Supplementary Material). 
         For the HM, as doping level $\delta$ increases, the excitation gap at $q=\pi$ becomes larger, while the $\omega=0$ peak's momentum position $q_0 = (1-\delta)\pi$ becomes smaller. For the HM+$V$, the excitation gap at $q=\pi$ has softened relative to the HM across all doping. For the HM+$g_{0}$+$g_{1}$, the extended {\it el-ph} couplings broadens the excitation peaks in the momentum range $(1-\delta)\pi \le q \le \pi$, with noticeable spectral weight extending to low frequencies (or even zero at small doping). 
    }
    \label{pic:L40_sqw}
\end{figure*}

To solve for the ground states of these models, we use the density matrix renormalization group (DMRG)~\cite{whiteDensityMatrixFormulation1992, whiteDensitymatrixAlgorithmsQuantum1993a} method; and for the HM+$g_{0}$+$g_{1}$ we employ a local basis optimization (LBO)~\cite{zhangDensityMatrixApproach1998a} for the phonon degrees of freedom in Eq.~\ref{eq:hhm}. 
We then use time-dependent DMRG (tDMRG)~\cite{vidalEfficientClassicalSimulation2003, whiteRealTimeEvolutionUsing2004, paeckelTimeevolutionMethodsMatrixproduct2019}, with a dynamical LBO~\cite{brocktMatrixproductstateMethodDynamical2015a} for the HM+$g_{0}$+$g_{1}$, to calculate the  correlator
\begin{equation}
S(r,t) = \left<\hat{s}^z_{L/2+r}(t)\hat{s}^z_{L/2}(0)\right> - \left<\hat{s}^z_{L/2+r}(0)\hat{s}^z_{L/2}(0)\right>,
\label{eq:srt}
\end{equation}
which is measured at uniformly spaced discrete time steps $t_n=n\delta t$, $n=0, 1,\cdots, N$, until reaching the maximum time $T=N\delta t$. $S(q,\omega)$ is then obtained by Fourier transform. To reduce the effect of the cutoff at finite time $T$, a decaying window function $\mathcal{W}_\sigma(t)$, where $\mathcal{W}_\sigma(T) \approx 0$, is multiplied to the time signal. The resulting spectra is
\begin{eqnarray}
    S_{\sigma}(q,\omega) & = & \frac{1}{L}\sum_r e^{-iq r}\int_{-\infty}^{\infty}e^{i\omega t}S(r,t)\mathcal{W}_\sigma(t) dt, \\ \nonumber
    & \approx & \frac{\delta t}{L}\sum_r\sum_n 2\Re(e^{i(\omega t_n-qr)}S(r,t_n))\mathcal{W}_\sigma(t) dt.
    \label{eq:sqw}
\end{eqnarray}
We use a window function $\mathcal{W}_\sigma(t) = e^{-\frac{\sigma^2}{2}t^2}$, whose Fourier transformation $\mathcal{F}\left[\mathcal{W}_\sigma(t)\right]$ is a Gaussian with standard deviation $\sigma$. This window function broadens the spectra
\begin{equation}
    S_\sigma(q,\omega) = S(q,\omega) * \mathcal{F}\left[\mathcal{W}_\sigma(t)\right].
\end{equation}

We compute and compare $S(q,w)$ at various doping levels $\delta$ for the three models: HM, HM+$V$ and HM+$g_{0}$+$g_{1}$. 
Unless otherwise stated, the spectra have been evaluated for parameters $U=8t_h$ and $V=-t_h$ for the HM+$V$ or $\omega_0=0.2t_h$, $g_0=0.3t_h$, and $g_1=0.15t_h$ for the HM+$g_{0}$+$g_{1}$. These parameters have been shown to well-reproduce features in the experimental single-particle spectra as measured by ARPES~\cite{chenAnomalouslyStrongNearneighbor2021, tangTracesElectronphonon2023}. 

We first show results for relatively long chains with $L=80$ and relatively long total times to produce high resolution spectra with small finite-size effects to elucidate the spectral changes for representative doping $\delta=20\%$. As shown in Figs.~\ref{pic:L80_sqw}(a) and (b), both the simple HM and HM+$V$ have gapless spectra at momentum $q_0 = (1-\delta)\pi$, and with increasing momentum from $q_0$ to $\pi$, the excitation gap grows monotonically. Similar to the overall renormalization of the single-particle bandwidth and Fermi velocity with the introduction of the extended Hubbard interaction~\cite{chenAnomalouslyStrongNearneighbor2021,tangTracesElectronphonon2023}, the spectra for the HM+$V$ [shown in panel (b)] between $q_0$ to $\pi$ are noticeably softer than the spectra of the simple HM. In other words, a spectral gap at $q=\pi$ extends to higher energies in the absence of extended interaction $V$.

The introduction of extended {\it el-ph} coupling produces qualitative differences in $S(q,\omega)$ compared to both the HM and HM+$V$. For the HM+$g_{0}$+$g_{1}$, shown in Fig.~\ref{pic:L80_sqw}(c), the lowest excitation peak becomes significantly broadened across momenta $q\in(q_0,\pi)$, with spectral weight extending to very low energy (approaching $\omega=0$). As a comparison of the three models, we plot the energy distribution curves (EDC) at $q=\pi$ in Fig.~\ref{pic:L80_sqw}(d), highlighting both the softening of the gap in the HM+$V$ and the significant broadening and low energy spectral weight caused by the extended {\it el-ph} couplings.

An additional, but not so obvious, difference induced by the extended {\it el-ph} coupling occurs in the apparent ``$2k_F$'' position. We show the momentum distribution curves (MDC) at $\omega=0$ in Fig.~\ref{pic:L80_sqw}(e). For both the HM and HM+$V$, the spectral peak position is well defined at $q_{0} = (1-\delta)\pi$. The introduction of extended {\it el-ph} coupling broadens and shifts the peak towards $\pi$.

Due to the numerical complexities associated with the addition of phonons, we perform the remaining simulations on short 40-site chains. The resulting spectral resolution is sufficient to demonstrate the impact of extended interactions on $S(q,\omega)$ as a function of hole doping $\delta$, and the differences between an effective attractive interaction $V$ and extended {\it el-ph} coupling. 

\begin{figure}[b!]
    \begin{center}
        \includegraphics[width=\picWidth]{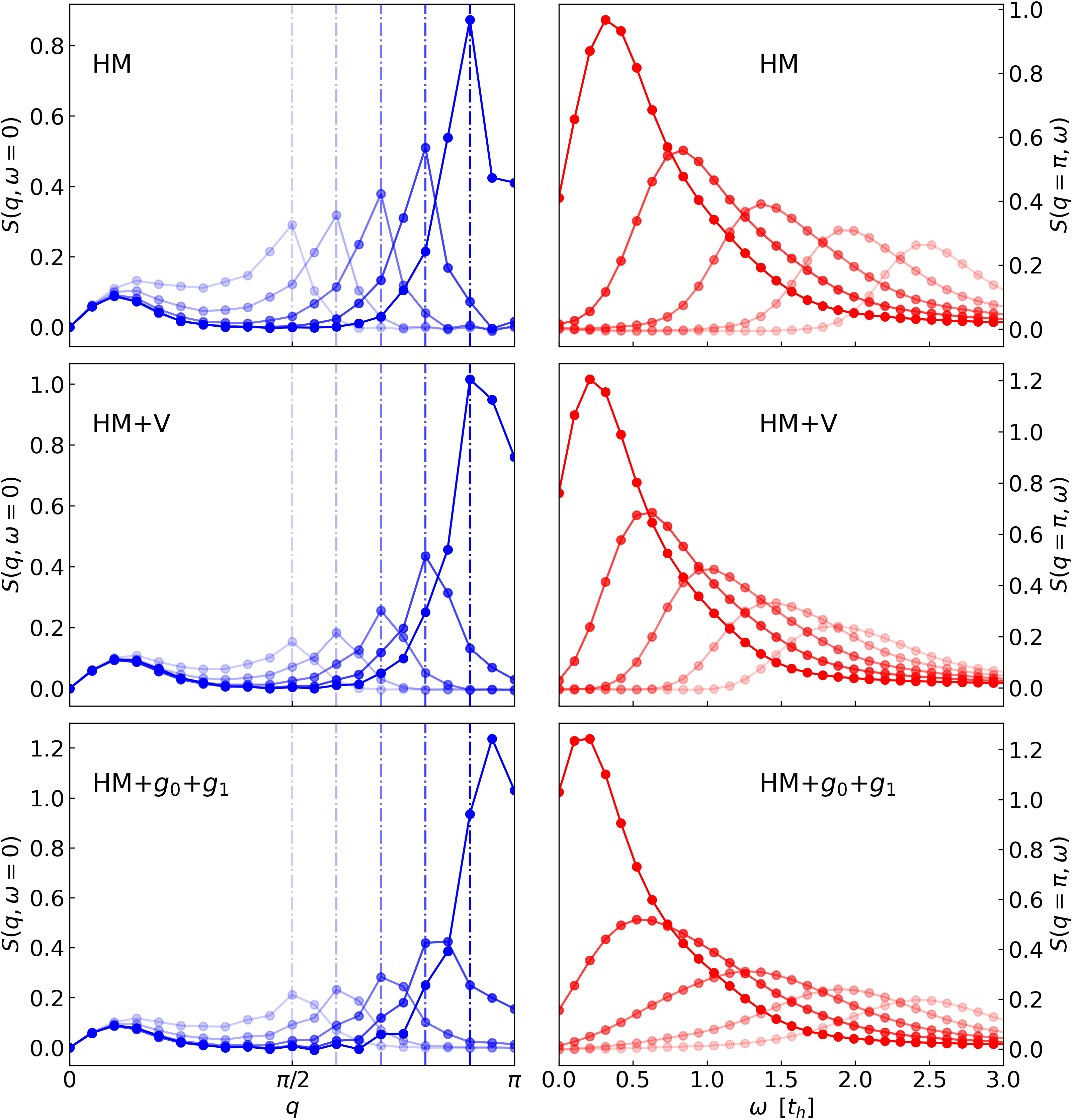} 
    \end{center}
    \caption{
        \textbf{MDC and EDC vs models.} 
         MDCs and EDCs for different models at different doping from $10\%$ to $50\%$ on the 40-site chain. Darker colors correspond to results at lower doping.
         Left column: MDCs at $\omega=0$. For both the HM and HM+$V$, the gapless ``$2k_F$'' peak appears at $q_{0} = (1-\delta)\pi$, marked by the dashed vertical lines. At lower doping, this peak shifts away from $q_{0}$ toward $q = \pi$ with the addition of extended {\it el-ph} coupling. Right column: EDCs at $q=\pi$. The HM+$V$ has lower peak positions compared to the HM across all doping, but with similar peak widths. Compared to the HM and HM+$V$, the features in the HM+$g_{0}$+$g_{1}$ are broader and the spectral weight extends to lower energy, even at relatively high doping levels. The peak positions themselves, compared to the HM, have lower energy at low doping ($\delta < 30\%$), but this becomes more comparable at high doping.
    }
    \label{pic:mdc_edc}
\end{figure}

In Fig.~\ref{pic:L40_sqw}, we show $S(q,\omega)$ for different models (rows) with increasing doping $\delta$ (columns). The time-dependent spin-spin correlators in real-space $S(r,t)$ can be found in Fig.~\ref{pic:L40_srt} of the Supplementary Material. Even with the lower resolution afforded on the 40-site chain, one still sees that both the HM and HM+$V$ at doping level $\delta$ possess gapless $S(q,\omega)$ spectra at momentum $q_0 = (1-\delta)\pi$. In each case, for momenta from $q_0$ to $\pi$, this excitation gap grows monotonically. As doping $\delta$ increases, $q_0$ becomes smaller and the gap at $\pi$ becomes larger (see Fig.~\ref{pic:L40_sqw} rows A and B). Compared to the simple HM, the HM+$V$ has a smaller excitation gap at $q=\pi$ across all doping levels, as also seen in the higher resolution spectra of Fig.~\ref{pic:L80_sqw}.

For the HM+$g_{0}$+$g_{1}$ with $\delta>0$ (Fig.~\ref{pic:L40_sqw} row C), the lowest excitation peak is significantly broadened across momentum $q\in(q_0,\pi)$ at all doping levels, and there is noticeable spectral weight extending to very low energy (even approaching $\omega=0$ at low doping). 

As a comparison of the three models, we plot the energy distribution curves (EDC) at $q=\pi$ in the right column of Fig.~\ref{pic:mdc_edc}, demonstrating broadened peaks and low energy spectral weight caused by the extended {\it el-ph} couplings. In terms of the peak positions at $q=\pi$, compared to the HM, the HM+$g_{0}$+$g_{1}$ has lower energy peaks at low doping ($\delta<30\%$), but the peak positions at higher doping are more comparable; however, at all doping levels the spectra of the HM+$g_{0}$+$g_{1}$ are significantly broadened compared to both the HM and HM+$V$. We also show the momentum distribution curves (MDC) at $\omega=0$ obtained from Fig.~\ref{pic:L40_sqw} in the left column of Fig.~\ref{pic:mdc_edc}. For both the HM and HM+$V$, the spectral peaks are well defined at $q_{0} = (1-\delta)\pi$. With extended {\it el-ph} coupling, the peaks broaden and shift toward $\pi$ at small doping, as observed with higher resolution on the 80-site chain. 

We have shown that the extended {\it el-ph} coupling induces significant changes to the spin dynamics across a wide range of doping levels. Specifically, without the extended {\it el-ph} coupling, $S(q,\omega)$ is gapless at momentum $q_0=2k_F=(1-\delta)\pi$ and has a gap that widens from $q_0$ to $\pi$. The extended {\it el-ph} coupling significantly broadens the excitations across this momentum range and introduces noticeable spectral weight down to zero energy for doping levels up to at least 50\%, as examined here. Compared to HM, the peak positions at $q=\pi$ are softened at smaller doping but remain comparable, yet broad, at higher doping levels. There is also a shift of $q_0$ from $2k_F$ towards $\pi$ caused by the extended {\it el-ph} coupling. We also have demonstrated that an effective attractive interaction $V$ can not emulate well all these quantitative, or even qualitative changes in $S(q,\omega)$ induced by the extended {\it el-ph} coupling.

Compared to the enhancement of the holon folding branch seen in the single particle spectral function $A(q,\omega)$, which is relatively weak and disappears quickly as the doping level increases,~\cite{chenAnomalouslyStrongNearneighbor2021,tangTracesElectronphonon2023} the impact of {\it el-ph} coupling can be more readily observed in $S(q,\omega)$ across a wide range of doping due to the modification of spectral weight. The position of $q_0$, the peak positions at $q=\pi$, and the general shape of both MDC and EDC cuts at lower doping can provide vital clues about the strength of extended {\it el-ph} coupling in these materials. Our work points out that a promising and perhaps more sensitive experimental assessment of extended {\it el-ph} coupling in the doped 1D cuprates can come, for example, from measurements of $S(q,\omega)$ by resonant inelastic X-ray scattering (RIXS)~\cite{amentResonantInelasticXray2011a}. 
 
\begin{acknowledgements}

We are grateful for detailed discussions with Jiarui Li, Wei-Sheng Lee, Z.-X. Shen, and Yao Wang. This work was supported by the U.S. Department of Energy, Office of Basic Energy Sciences, Division of Materials Sciences and Engineering, under Contract No.~DE-AC02-76SF00515. The computational results utilized the resources of the National Energy Research Scientific Computing Center (NERSC) supported by the U.S. Department of Energy, Office of Science, under Contract No.~DE-AC02-05CH11231. Some of the computing for this project was performed on the Sherlock cluster. We would like to thank Stanford University and the Stanford Research Computing Center for providing computational resources and support that contributed to these research results.
\end{acknowledgements}

\bibliography{reference}

\appendix
\setcounter{equation}{0}
\setcounter{figure}{0}
\setcounter{table}{0}
\makeatletter
\renewcommand{\theequation}{S\arabic{equation}}
\renewcommand{\thefigure}{S\arabic{figure}}

\section*{Supplementary Material}

\subsection*{DMRG details for L=80}

\begin{figure*}[th]
    \begin{center}
        \includegraphics[width=\widePicWidth]{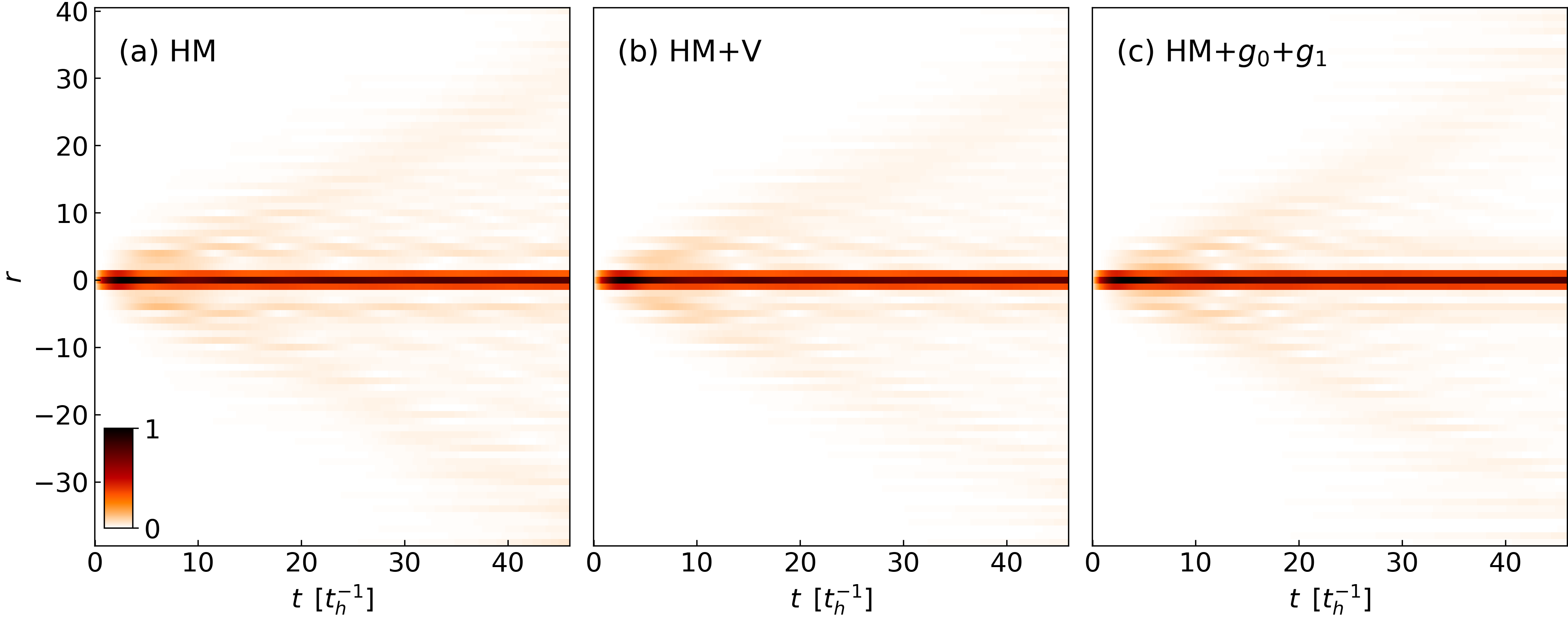} 
    \end{center}
    \caption{
         \textbf{Time-dependent spin-spin correlations in real space on a chain with L=80.}
         $\left|S(r,t)\right|$ (see Eq.\ref{eq:srt}) for different models at 20\% dopings on 80-site chains. (a) Hubbard model (HM), (b) extended Hubbard model (HM+$V$), and (c) Hubbard-extended Holstein model (HM+$g_{0}$+$g_{1}$). Each plot is normalized to its own maximum. 
    }
    \label{pic:L80_srt}
\end{figure*}

In the DMRG algorithm, we keep up to $m=800$ states for all models and a bare phonon basis $n=20$ and an optimal LBO phonon basis $n_{o}=3$ for the HM+$g_{0}$+$g_{1}$. 
To find the ground state, we perform up to 36 DMRG sweeps and then perform up to $900$ time evolution steps in tDMRG with time interval $\delta t=0.05t_h^{-1}$. For the simple HM, the DMRG ground state truncation error was around $1\times10^{-8}$ and the maximum truncation error for time evolution was around $1.4\times10^{-6}$. For the HM+$V$, the DMRG ground state truncation error was around $1.3\times10^{-8}$ and the maximum truncation error for time evolution was around $2.1\times10^{-6}$.  
Finally, for the HM+$g_{0}$+$g_{1}$, the DMRG ground state truncation error with respect to $m$ was around $6.8\times10^{-7}$, with an LBO phonon basis truncation error around $4.3\times10^{-6}$; and for time evolution of the HM+$g_{0}$+$g_{1}$, the maximum truncation error with respect to $m$ was around $1.1\times10^{-5}$, with a maximum truncation error of the dynamical LBO basis around $7.9\times10^{-5}$.
\newline

\subsection*{DMRG details for L=40}

\begin{figure*}[th]
    \begin{center}
        \includegraphics[width=\widePicWidth]{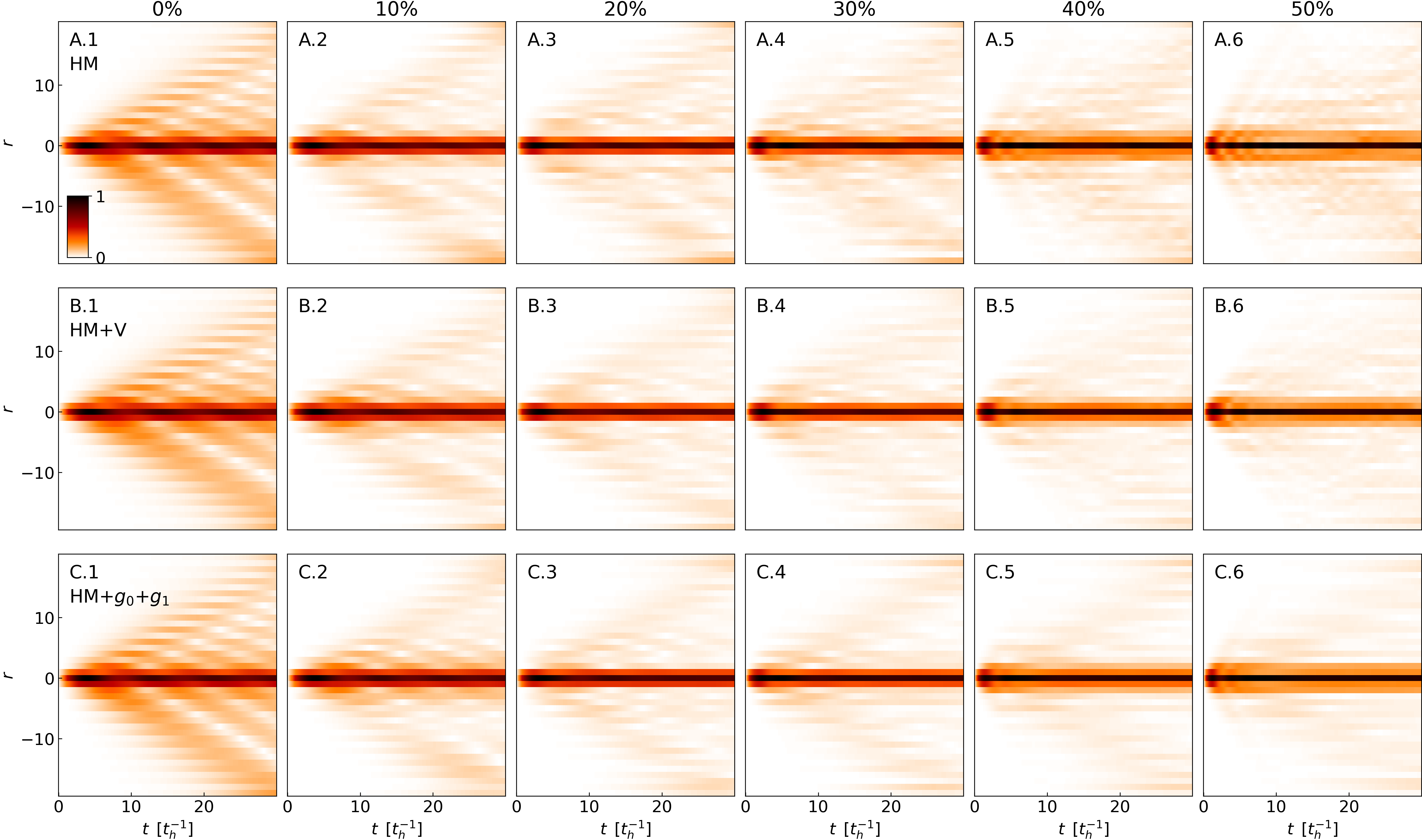} 
    \end{center}
    \caption{
         \textbf{Time-dependent spin-spin correlation in real space.}
         $\left|S(r,t) \right|$ (see Eq.\ref{eq:srt}) from different models at different dopings on 40-site chains. Row A is for Hubbard model (HM), row B is for extended Hubbard model (HM+$V$), and row C is for Hubbard-extended Holstein model (HM+$g_{0}$+$g_{1}$). Each plot is normalized to its own maximum. 
    }
    \label{pic:L40_srt}
\end{figure*}

For the 40-site chain, we kept a number of DMRG block basis states up to $m=800$ for all models, with a number of bare phonons up to $n=20$ and the number of optimal LBO phonons equal to $n_o=3$ for the HM+$g_{0}$+$g_{1}$. We perform up to 30 DMRG sweeps to determine the ground state and then perform up to $600$ time evolution steps with time interval $\delta t=0.05t_h^{-1}$. For a doping level $\delta=20\%$ as an example, we list the truncation errors in our simulations. For the HM, the ground state truncation error was around $3\times10^{-10}$, and the maximum truncation error for time evolution was around $1\times10^{-6}$. For the HM+$V$, the ground state truncation error was around $3\times10^{-10}$, and the maximum truncation error for the time evolution was around $2\times10^{-6}$. For the HM+$g_{0}$+$g_{1}$, the ground state final sweep truncation error with respect to $m$ was around $1\times10^{-7}$ and the LBO phonon basis truncation error was around $4\times10^{-6}$. During time evolution with the dynamical LBO, the maximum truncation error with respect to $m$ was around $1\times10^{-5}$ and the maximum truncation error of the phonon basis was around $2\times10^{-5}$. The truncation errors at other doping levels are close to these values (within a factor of $\sim 2$ to 3).

\subsection*{Dependence on $g_1$}
\begin{figure*}[th]
    \begin{center}
        \includegraphics[width=\widePicWidth]{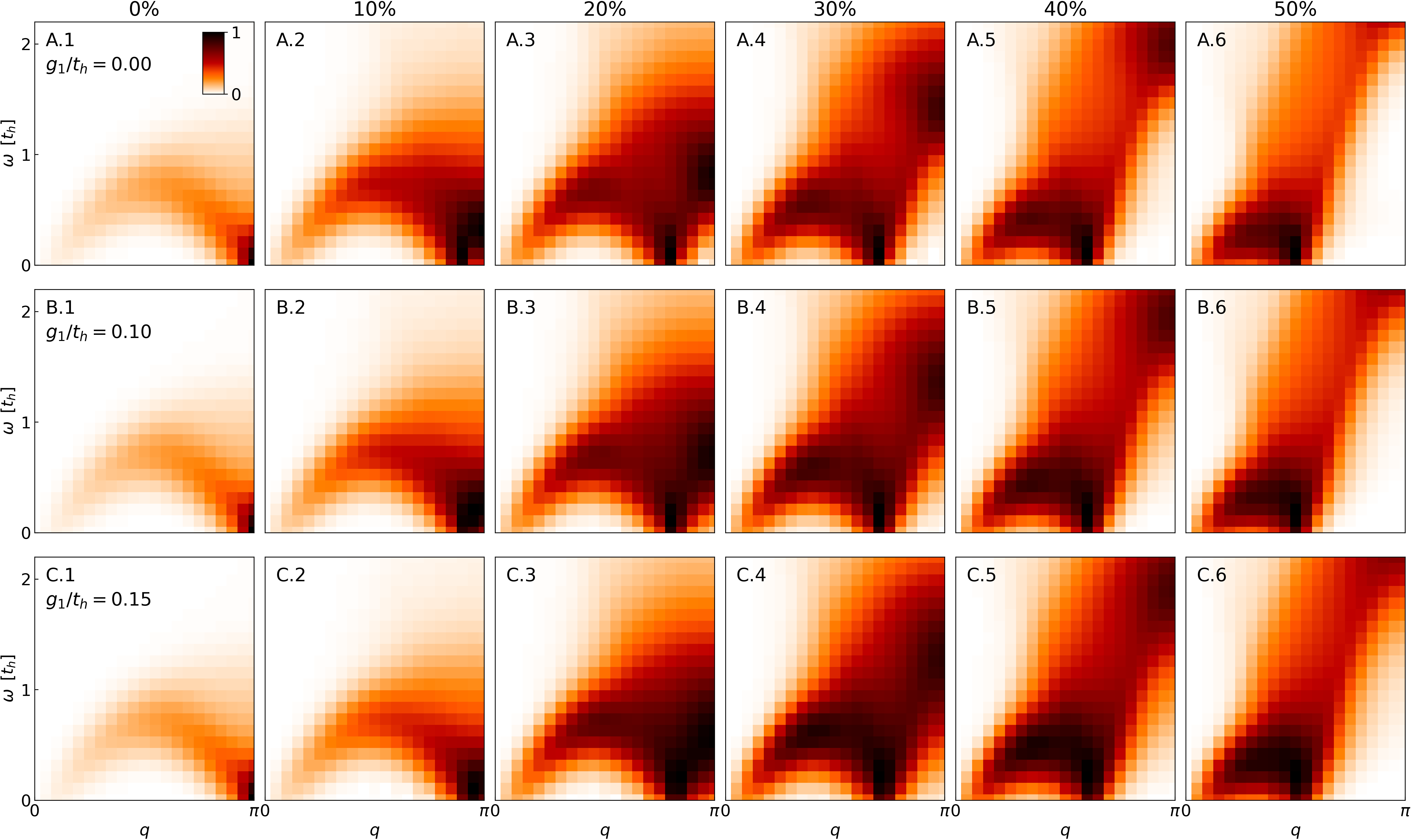} 
    \end{center}
    \caption{
         \textbf{$S(q,\omega)$ for different values of the extended \textit{el-ph} interaction $g_1$}.
    }
    \label{pic:L40_sqw_g1}
\end{figure*}

\begin{figure*}[th]
    \begin{center}
        \includegraphics[width=\picWidth]{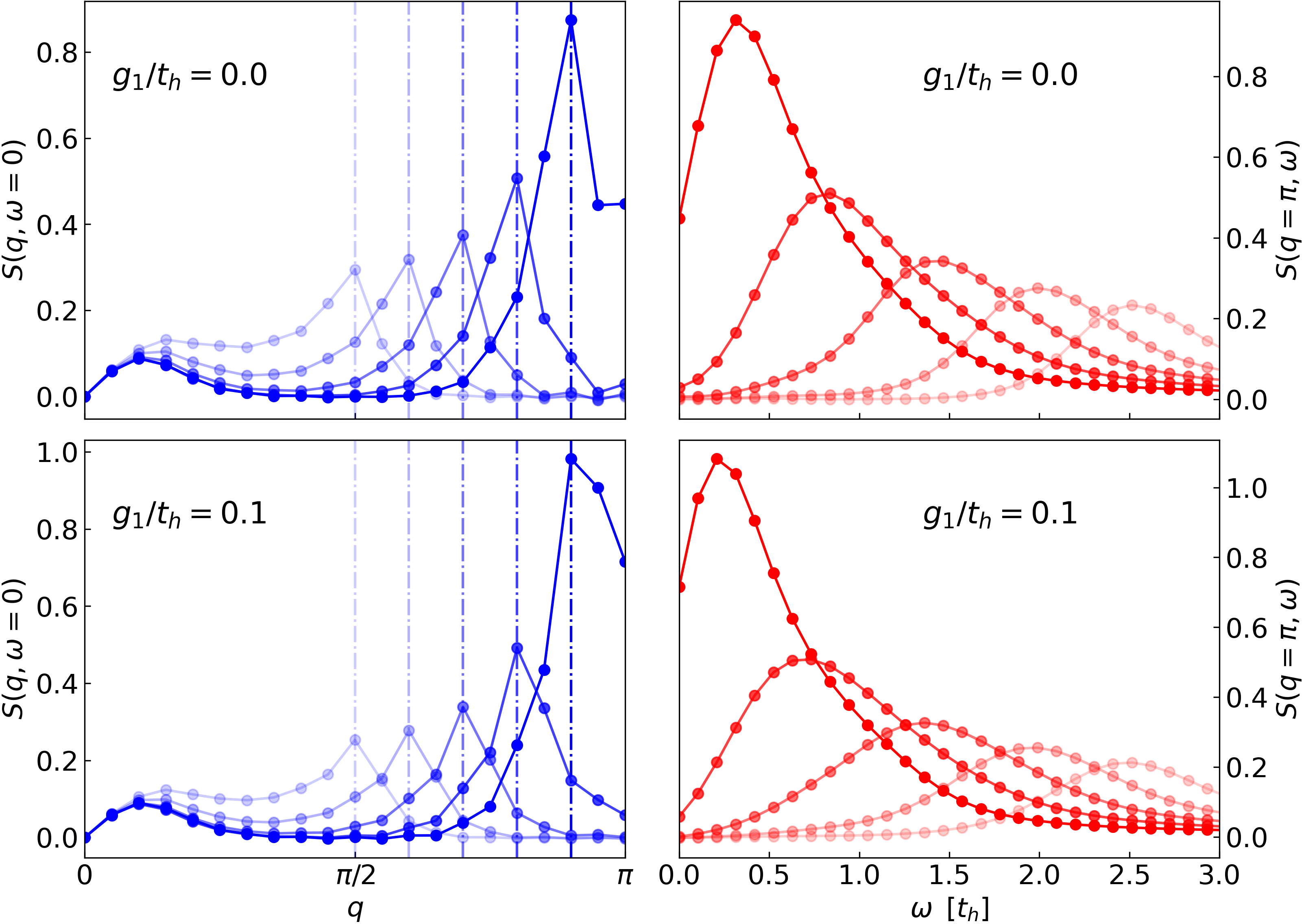} 
    \end{center}
    \caption{
        \textbf{MDC and EDC vs extended {\it el-ph} coupling strength \bm{$g_1$}.} 
         Doping from $10\%$ to $50\%$. Darker color indicates lower doping. Left column: MDCs at $\omega=0$. Right column: EDCs at $q=\pi$. For $g_1=0$, both the EDCs and MDCs are similar to those obtained in the HM. Evident changes associated with increasing $g_1$ include a shift of $q_0$ in the MDCs, broadening of peaks in the EDCs, and a softening of EDC peaks at lower doping levels. 
    }
    \label{pic:mdc_edc_g1}
\end{figure*}

\begin{figure*}[th]
    \begin{center}
        \includegraphics[width=\widePicWidth]{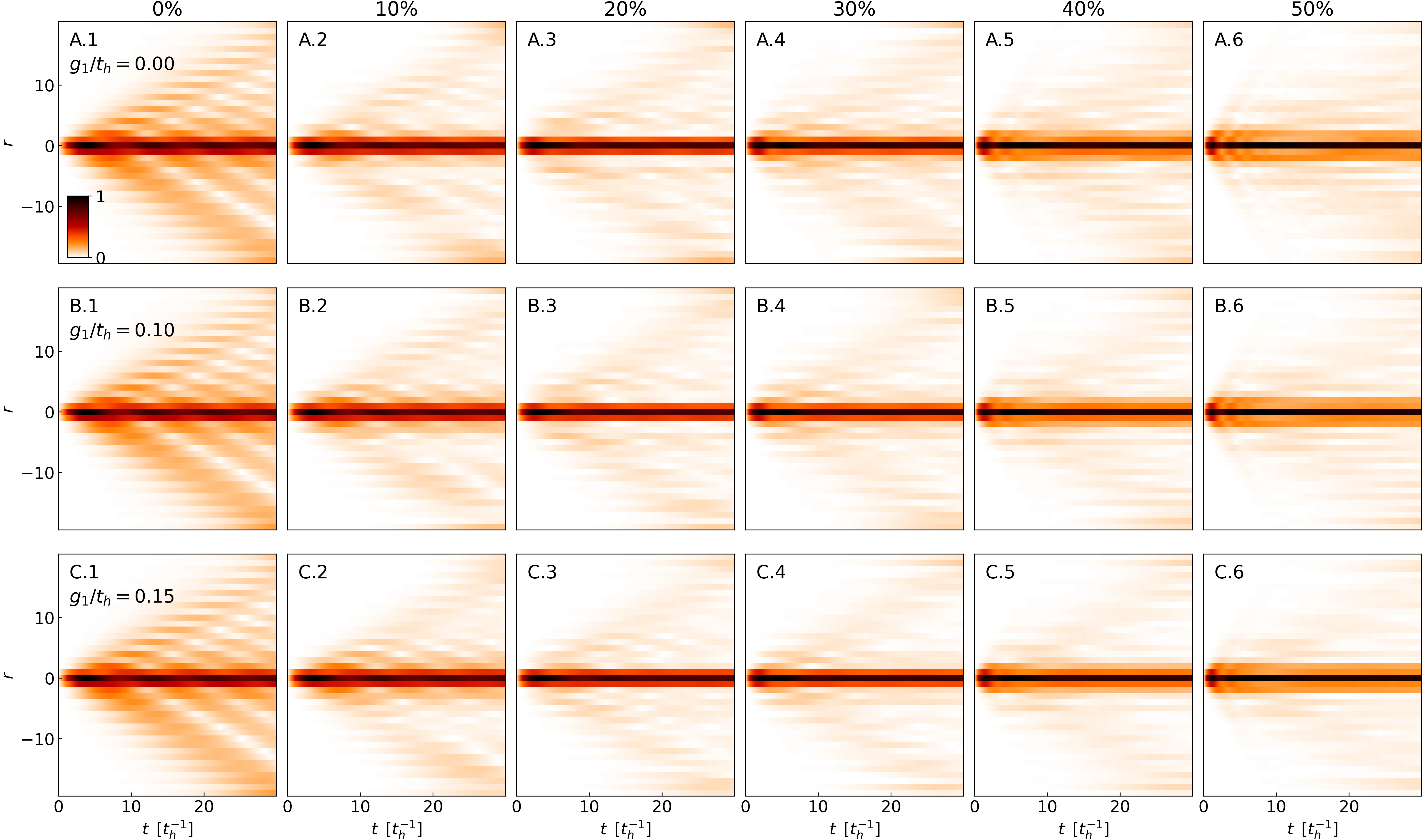} 
    \end{center}
    \caption{
          \textbf{$S(r,t)$ for different values of the extended \textit{el-ph} interaction $g_1$}.
    }
    \label{pic:L40_srt_g1}
\end{figure*}

To demonstrate the spectral dependence on the extended {\it el-ph} coupling, we plot full spectra as a function of doping in Fig.~\ref{pic:L40_sqw_g1}, and then extract MDCs at $\omega=0$ and EDCs at $q=\pi$ in Fig.~\ref{pic:mdc_edc_g1}, for different values of $g_1$. For $g_1=0$, the MDCs and EDCs are similar to the results of the HM, as the local Holstein coupling primarily renormalizes the effective on-site Hubbard interaction. One sees the previously discussed modifications to the spectra that appear as one increases $g_1$ to $0.1 t_h$ in the second row of Fig.~\ref{pic:L40_sqw_g1} and further increases $g_1$ to $0.15 t_h$ in the third row, as in Fig.~\ref{pic:L40_sqw} of the main text. Corresponding real-space and time-dependent spectra can be found in Fig.~\ref{pic:L40_srt_g1}. 

\end{document}